\documentclass[]{aastex631}
\usepackage[utf8]{inputenc}
\usepackage{nameref}
\usepackage{wasysym}
\usepackage{makecell}

\usepackage{rotating}
\usepackage{graphicx}
\usepackage{sidecap}
\usepackage{soul}
\usepackage{comment}
\usepackage[title]{appendix}

\usepackage{BibDef}
\usepackage{natbib}

\graphicspath{{img/}}

\begin{document}
	
\title{Possible {\it in situ} formation of Uranus and Neptune via Pebble Accretion}

\author{Claudio Valletta}
\affiliation{Center for Theoretical Astrophysics and Cosmology \\
			Institute for Computational Science, University of Zürich \\
			Winterthurerstrasse 190, 8057 Zürich, Switzerland}

\author{Ravit Helled}
\affiliation{Center for Theoretical Astrophysics and Cosmology \\
			Institute for Computational Science, University of Zürich \\
			Winterthurerstrasse 190, 8057 Zürich, Switzerland}

\accepted{for publication in ApJ}

\begin{abstract}
\large
The origin of Uranus and Neptune is still unknown. In particular, it has been challenging for planet formation models to form the planets in their current radial distances within the expected lifetime of the solar nebula.  
In this paper, we simulate the \textit{in-situ} formation of Uranus and Neptune via pebble accretion and show that both planets can form within $\sim$ 3 Myr at their current locations, and have final compositions that are consistent with the heavy-element to H-He ratios predicted by structure models. 
We find that Uranus and Neptune could have been formed at their current locations. In several cases a few earth masses (M$_\oplus$) of heavy elements are missing, suggesting that Uranus and/or Neptune may have accreted $\sim$1 -- 3 M$_\oplus$ of heavy elements after their formation via planetesimal accretion and/or giant impacts.
\end{abstract}

\large
\section{Introduction} \label{sec:introduction}

Uranus and Neptune are the outermost planets in the solar system. They have masses of $14.5$ and $17.1$ M$_\oplus$ and are located at $19.1$ and $30$ AU, respectively.  Unlike Jupiter and Saturn, the hydrogen-helium (hereafter, H-He) envelopes of Uranus and Neptune are relatively small fractions of their total masses \citep{Fortney2010,Nettelman2013,Helled10,Helled19}.

The {\it in situ} formation of Uranus and Neptune has been a challenge for a few decades because at such large orbital locations the formation time-scale typically exceeds the expected lifetimes of protoplanetary disks \citep{Safronov1969}. A potential solution for these long formation timescales is to assume that Uranus and Neptune formed closer to the Sun and then migrated outwards
\citep{thommes99,tsiganis05}, however there is no evidence that Uranus and Neptune indeed formed farther in.

An important constraint for formation models of Uranus and Neptune is to reproduce their solid-to-gas ratio as inferred by structure models. 
For example, it was shown by \cite{Helled14} when using high heavy-element accretion rates as suggested by \cite{Rafikov2011}, the problem in forming Uranus and Neptune is no longer the formation time-scale but rather finding an agreement with the right heavy-element to H-He ratio, $M_Z/M_{H-He}$.
In addition, it is difficult to explain why Uranus and Neptune accreted H-He without undergoing runaway accretion of gas becoming gas giants similar to Saturn and Jupiter.
A very specific set of parameters is required to reproduce the observed mass and composition of Uranus and Neptune. This is known as the \textit{fine tuning} problem.

The formation at large radial distances was presented by \cite{Goldreich04}. It was shown that Uranus and Neptune could form at their present-day orbital locations if the solid surface density is high enough (a few times that of the minimum mass solar nebula) and the solids have small sizes (a few cm). Such small solids can be generated by planetesimal collisions.  
\par 

Further studies on planet formation via pebble accretion have confirmed that the pebble accretion rate can be high enough to form giant planets at large orbital locations \citep[e.g.,][]{Lambrechts14,Bitsch15}. However, in these studies the H-He  mass  during the planetary growth was often assumed to be equal to 10\% of the total planetary mass.
The formation of Uranus and Neptune via pebble accretion was presented in \cite{lambrechts15}. In this study the dissolution of the accreted pebbles within the envelope was modeled assuming that the icy component of pebbles is uniformly sublimated into the envelope while the refractory material joins the  core. In this model the envelope's composition is uniform at all time, but enriched with respect to the stellar value. 
\cite{lambrechts15} showed that Uranus and Neptune can be formed in their current locations and that the inferred $M_Z/M_{H-He}$ agrees with the ones inferred from structure models and is rather insensitive to the assumed orbital's location. 
\par

An investigation of the formation of mini-Neptunes via pebble accretion was presented by \cite{Venturini17}. Heavy-element enrichment of the planetary envelope was included in the equation of state (EOS) and opacity calculations.
The authors consider both pebble and planetesimal accretion and explore the effect of the envelope's opacity on the planetary growth. The occurrence rate of mini-Neptunes was found to be between 10 to 80 \% when assuming pebble accretion. 
In the context of the formation of Uranus and Neptune, it was shown that pebbles accretion combined with a high opacity allow the formation of mini-Neptunes at 30 AU. On the other hand, accretion of planetesimal was found to lead to the formation of mini-Neptunes at small radial distances ($\sim$ 5 AU) with the occurrence rate depending on the envelope's opacity.

In both of the studies of \cite{lambrechts15} and  \cite{Venturini17} the heavy-element enrichment of the envelope was not modeled self-consistently. 
The amount of pebbles dissolved in the envelope was not computed but rather {\it assumed}: half of the pebble's mass, represented by rocky material, was set to join the core while the other half composed of water was assumed to be uniformly mixed in the planetary envelope. This is clearly an over-simplification since also rocky pebbles are expected to completely dissolve in the envelope due to their small sizes once the core mass reaches $\sim$2 M$_\oplus$ \citep[e.g.,][]{Valletta19,Brouwers2017}. In addition, the heavy elements in the envelope of growing protoplanets are likely to have a gradual distribution and not being homogenously mixed \citep{Lozovsky2017, Helled17, Valletta19,Valletta20}.

In \cite{Valletta20} (hereafter VH20) we presented a numerical framework to model giant planet formation accounting  self-consistently the effect of heavy-element enrichment of the planetary envelope on the equation of state and opacity calculation. In addition,  we follow the deposition of the heavy elements in the envelope allowing for composition gradients, and calculate the consequent accretion of solids and gas. 

In this study, we apply the model presented in VH20 to investigate the \textit{in-situ} formation of Uranus and Neptune via pebble accretion as described in section~\ref{method}. 
We search for formation paths in which the formation timescale and final compositions are consistent with the lifetime of the solar nebula and the inferred compositions of Uranus and Neptune from structure models, respectively. Our results are presented in  section~\ref{Results}. We next discuss the connection to giant impacts on Uranus and Neptune 
in section~\ref{impacts}. Finally, our conclusions are summarized in section~\ref{conclusion}.

\section{Model}

\label{method}
We simulate the {\it in-situ} formation of Uranus and Neptune via pebble accretion using the formation model presented in VH20.
We adapted the numerical code MESA (Modules for experiment in stellar astrophysics) \citep{Paxton2010,Paxton2013,Paxton2015,Paxton2018,Paxton2019} to simulate the formation of giant planets. 
The simulations begin with a heavy-element core with a mass of $M_{icore} = 0.01 M_{\oplus}$ and a negligible envelope. The consequent planetary growth is insensitive to small changes in the initial conditions because at early stages gas accretion is  limited for slightly different choices of core and envelope masses \citep[e.g.,][]{POLLACK1996,ida04}
We use the EOS module developed by \cite{Muller20} to properly model a mixture of H-He and heavy elements (represented by H$_2$O) in planetary conditions. 
The total opacity is the sum of the gas and the dust opacity where the former is computed via the standard low temperature \citep{Freedman2014} Rosseland tables and the latter is determined using the prescription of \citet{Valencia2013}. 
The outer temperature and pressure are set to the disk's pressure and temperature \citep{Piso14} at the planet's location.
We compute the mass and energy deposition by the accreted pebbles and we include the presence of the the deposited H$_2$O in the EOS and opacity calculations. More details on the model can be found in VH20.

\subsection{Accretion rates}
The heavy-element accretion rate is a key factor when modeling the formation of Uranus and Neptune. As shown in \cite{Helled14} very high solid accretion rates result in a formation time-scale that is too short and the forming planet would then be a giant planet. On the other hand, low accretion rates lead to the formation of mini-Neptunes/super-Earths.
Accounting for the heavy-element enrichment of the envelope affects the gas accretion rate:  H-He is accreted more rapidly as a consequence of the higher mean molecular weight of the envelope \citep{stevenson82,Venturini16,Venturini17,Valletta20}. 

\noindent \underline{Pebble accretion rate:}\\
We use the pebble accretion rate derived by \cite{Lambrechts14} (their Eq.~31) where we assume $\beta = $ 500 g $cm^{-2}$; this parameter regulates the gas surface density.  
The planetary growth depends on $\beta$ and the disk's metallicity. 
In the paper we fix the gas surface density and investigate the sensitivity of our results to the disk metallicity and viscosity. We also determine whether the pebble accretion rate is governed by 2D or 3D effects  (using Eq.~5 of \citet{Venturini17}). 
The expressions for the heavy-element accretion rates implemented in this work are given in Appendix~\ref{saccrateapp}.
We follow \cite{Piso14} to determine the pressure and temperature at the planet's location.
The planetary growth depends on the solid surface density at the planet's location. In Appendix~\ref{diskmodelapp} we show that the predicted solid surface density at 20 AU of our disk's profile is similar to the minimum mass solar nebula model \citep{Hayashi1981}. Nevertheless, it should be kept in mind that the unknown solid surface density remains an important unknown property in planet formation models.



The accreted pebbles are assumed to be composed of pure water. In Appendix~\ref{composition} we investigate the sensitivity of the  results to the assumed composition of pebbles.
In order to mimic gradual gas dissipation we multiply $\beta$ with the exponential factor $\exp(-t/ 3 \; \mathrm{Myrs})$ as suggested by \cite{Lambrechts14}. 
The heavy-element accretion rate  strongly depends on the assumed  metallicity and viscosity of the disk but is only weakly affected by the orbital location 
In section~\ref{Results} and~\ref{impacts} we present selected successful runs that lead to Uranus-like and Neptune-like while in Appendix~\ref{parameters} we present results for a larger set of assumed parameters.

\noindent \underline{Gas accretion rate:}\\
We determine the gas (H-He) accretion rate numerically at each time-step by demanding that the computed planet's radius equals R$_A$ \citep{lissauer2009}:
\begin{equation}
\label{radius}
    R_A=\frac{GM_p}{c_s^2+4GM_p/R_H},
\end{equation}
where M$_p$ is the planet's mass, c$_s$ is the sound's speed, and R$_H$ is the Hill's radius. \\

\noindent \underline{Final composition:}\\
Interior structure models of Uranus and Neptune that fit their basic physical properties (mass, radius, rotation, gravity field) constrain $M_Z/M_{H-He}$ in Uranus and Neptune. 
The mass of H-He is between 1.25 and 3.53 M$_\oplus$ for Uranus and is between 1.64 and 4.15 M$_\oplus$ for Neptune, i.e. a minimum and maximum metallicity of 0.75 and 0.9 \citep{Nettelman13,Helled11}.
Planets like Uranus and Neptune are expected to form if the disk dissipates when the protoplanet has a mass of H-He compatible with observations. 
Our simulations are terminated when the planet has accreted $\sim$ 3.5 and 4.2 M$_\oplus$ of H-He when modelling the formation of Uranus and Neptune, respectively, since these values correspond to the upper bound of the H-He mass in Uranus and Neptune as determined by structure models \citep[e.g.,][]{Helled11}.

\noindent \underline{Formation likelihood:}

In order to assess whether our formation model can realistically explain the formation of Uranus and Neptune, we calculate the likelihood of our formation scenario by computing the probability that the disk dissipates when the planet has a H-He mass as predicted by  structure models as discussed above. 
The lifetime of the solar nebula is unknown,  therefore we assume a probability density function for the disk's lifetime $p(t)$.
The likelihood of forming Uranus and Neptune is then given by the integral of $p(t)$ in the time interval at which the planet has a H-He mass that is consistent with the mass range inferred by interior models and is given by:
\begin{equation}
\label{probability}
    f_{U,N} = \int_{t_{lower}}^{t_{upper}} p \; dt, 
\end{equation}
where $t_{lower}$ and $t_{upper}$ correspond to the time when the protoplanet has a mass equal to the lower and upper limit of the expected H-He in Uranus and Neptune. 
M$_{min}$ and M$_{max}$ are the minimum and maximum masses of H-He that are expected in Uranus and Neptune \citep{Helled2020}. We assume M$_{min-U}$=1.25 M$_\oplus$ and M$_{max-U}$ = 3.53 M$_\oplus$ for Uranus, and M$_{min-U}$=1.64 M$_\oplus$ and M$_{max-N}$= 4.15 M$_\oplus$ for Neptune. 
The formation likelihood depends on M$_{min-U(N)}$ and M$_{max-U(N)}$ and in Appendix~\ref{internalcomposition} we investigate the sensitivity of the results to these assumed values.

The likelihood of forming Uranus and Neptune as computed via Eq.~\ref{probability} provides  a guidance to determine the validity of a given model. The planetary growth depends on a set of parameters that are chosen ad-hoc to reproduce the formation of a Uranus- or Neptune- like  planet within the assumed disk lifetimes. 
A more advanced calculation of the formation likelihood should also include the probability of the different disk's parameters values (e.g., viscosity, metallicity). 

It is difficult to estimate the probability density function $p(t)$. We therefore consider  two different assumptions: (i) we assume that the disk's dissipation time is given by a gaussian function with a median value of 3 Myrs and a standard deviation of 1 Myrs; (ii) we use Eq.~2 of \cite{Venturini17} where $p(t)$ is given by an exponential function $e^{-t/\tau}$, with $\tau$ = 2.5 Myrs \citep[e.g.,][]{Ribas15,Pfalzner14}.

\section{Results}

\label{Results}
\begin{figure}[h]
	\centering
	\includegraphics[width=0.5\linewidth]{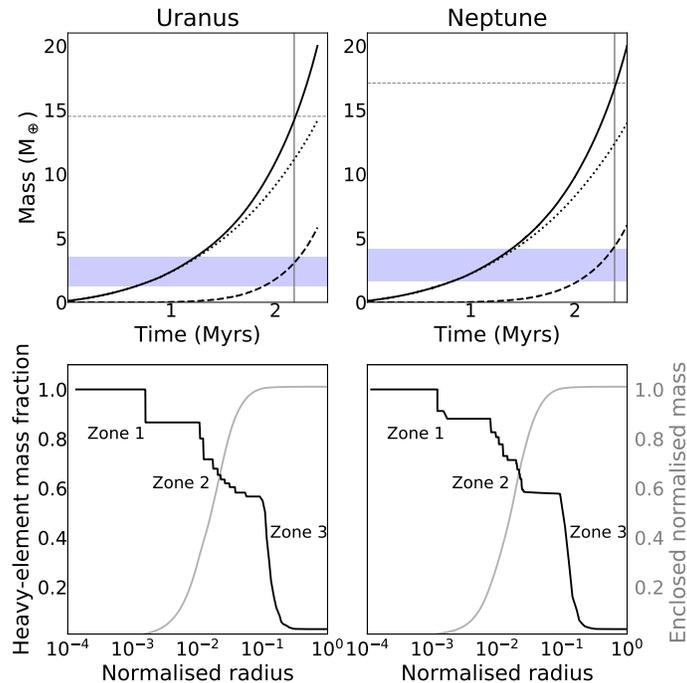}
	\caption{
	The formation of Uranus (left) and Neptune (right). 
	\textbf{Top panels:}
	Planetary mass as a function of time. The solid, dashed, and dotted lines represent the total planetary mass, the H-He mass, and the mass of heavy elements, respectively. 
	The grey horizontal lines indicate the mass of Uranus and Neptune, while the vertical grey line shows the required disk's dissipation time.
	The blue region indicates the H-He in Uranus and Neptune as inferred from interior models. 
    \textbf{Bottom panels:} internal profile when the disk dissipates. The black and grey line represent the heavy-element mass fraction
    and the enclosed mass as a function of the normalised radius, respectively.
	}
	\label{Uranus-Neptune}
\end{figure}

The top panels of Figure~\ref{Uranus-Neptune} show successful simulations of the formation of Uranus and Neptune for a given set of parameters. 
In both cases the disk's metallicity is assumed to be 1.45 Z$_\odot$, however, the disk's viscosity is 10$^{-3}$ and 5 $\times 10^{-4}$ for Uranus and Neptune, respectively. 
In Appendix~\ref{parameters} we investigate the formation of Uranus and Neptune considering a larger parameter space of assumed properties.  

The blue shaded region indicates the expected H-He mass as inferred from structure models. 
We find that at 20 AU, when the total mass is 14.5 M$_\oplus$ the planet has accreted 3 M$_\oplus$ of H-He, in agreement with Uranus' structure models. This is also the case for Neptune where the planet has accreted 4 M$_\oplus$ when the total mass is 17.1 M$_\oplus$.

In order to prevent Uranus and Neptune from becoming gas giants, the disk should dissipate by the time they reach their final masses. For the cases presented in Fig.~\ref{Uranus-Neptune}, we find that the dissipation time of the solar nebula should be $\sim$ 2.5 Myrs, compatible with the observed lifetimes of proto-planetary disks \citep{Ribas15,Pfalzner14}. 
The likelihood to form Uranus and Neptune according to the definition provided in Eq.~\ref{probability} is found to be 9.7\% and 10.6\%, respectively. 

It should be noted that we do not find a given set of model parameters that can reproduce the formation of both Uranus and Neptune exactly.
As we show below (Appendix~\ref{parameters}) when we assume the same parameters for the solar nebula, we typically find that at least  one of the planets is missing a couple of Earth masses of heavy elements and therefore must accrete additional heavy-elements or undergo a giant impact. 
While this is a plausible scenario as discussed in detail in section 3.1, it is also possible that the local conditions at 20 and 30 AU were different due to the $CO_2$ ice line that is expected to be between Uranus and Neptune \citep{Powell17}. This in return can that altered the the composition of the accreted pebbles and the gas metallicity.  

Figure~\ref{Uranus-Neptune} shows selected examples of simulations  that successfully lead to the {\it in-situ} formation of Uranus and Neptune. In appendix~\ref{parameters} we show the results for a larger set of assumed parameters that lead to the formation of Uranus-like and Neptune-like planets within the expected lifetime of the solar nebula.

The bottom panels of Figure~\ref{Uranus-Neptune} show the post-formation heavy-element distributions for Uranus and Neptune for these simulations. 
As discussed in VH20 the primordial heavy-element profile reflects the accretion rate of solids and gas during the planetary growth. 
It can be seen that the two planets have similar, but not identical, heavy-element profiles.
Both planets have composition gradients with the heavy-element mass fraction gradually decreasing towards the planetary surface; the plateau regions that are visible  in the bottom panels of Figure~\ref{Uranus-Neptune} are the result of convective mixing that leads to a homogeneous composition. 
Interestingly, the internal structures of the planets can be roughly divided into three main regions: (i) the central region with a high  heavy-element mass fraction of 0.8-1. The mass of the pure heavy-element compact core, is found to be about 4 and 3 M$_\oplus$ in Uranus and Neptune, respectively.
(ii) an envelope with a composition gradient where the heavy-element decreases with increasing radial distance. In this region the heavy elements are mixed with H-He and the heavy element mass fraction between 40-70\%. iii) an atmosphere dominated by H-He with a low heavy-element mass fraction of $\sim$0.03 , corresponding to 2.5 the solar value. 
Such three-zone division of the interior is qualitatively consistent with structure models of Uranus and Neptune \citep[e.g.,][]{Helled11, Nettelman13} and recent theoretical formation models \citep[e.g.,][]{Helled17, Valletta20,Ormel21}.

\subsection{Connection to giant impacts}

\label{impacts}
In some runs the planet's mass is equal to the mass of Uranus (Neptune) when the amount of H-He is about 3 M$_\oplus$ (4 M$_\oplus$).
However, in many cases the final planetary mass is less than the mass of Uranus (Neptune) when the desired mass of H-He is reached. For these cases, in order to reproduce  the masses  of Uranus and Neptune and their composition as inferred by structure models, the planets must accrete additional heavy elements after the disk's dissipation, typically of the order of a couple of M$_\oplus$.  
It should be noted, however, that the actual H-He mass in Uranus and Neptune could be lower of the order of $\sim$ 1.5 M$_\oplus$. If this is indeed the case, more massive impactors would be required. 
In Appendix~\ref{parameters} we show that this is a general trend for a relatively large set of parameters.

\begin{figure}[h]
	\centering
	
	\includegraphics[width=0.4\linewidth]{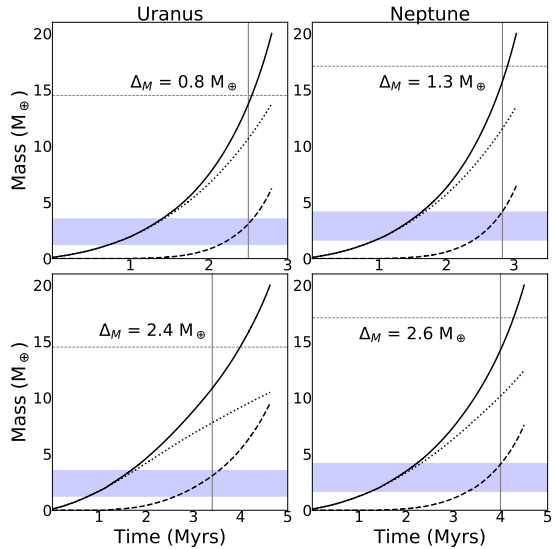}
		\caption{
		Same as Figure~\ref{Uranus-Neptune} but for different assumed parameters. In the top panels the disk's metallicity is set to 1.4 Z$_\odot$ while the disk's viscosity is 10$^{-3}$ and 5 $\times 10^{-4}$ for Uranus and Neptune, respectively. 
	In the bottom panels the disk's metallicity is 1.45 Z$_\odot$ with a disk viscosity is 1.5 $\times$ 10$^{-3}$ and 7.5 $\times 10^{-4}$ for Uranus and Neptune, respectively. 
	We also indicate by $\Delta_M$ the "missing heavy-element mass"  that should be accreted by the planet to reach the measured masses of Uranus (left panel plot) and Neptune (right panel plot).
	}
	\label{Uranus-Neptune2}
\end{figure}

Such post-formation enrichment, could occur as a result of giant impacts which seems to be needed to explain the observed properties of the planets \citep[e.g.,][]{Podolak2012,Reinhardt20,chau20}  or via late planetesimals accretion \citep{Shibata19,Shibata20}.
\cite{Reinhardt19} used SPH simulations and showed that the differences of the observed properties of Uranus and Neptune such as the axis tilt, the presence of moons, and the inferred moment of inertia, could be a result of giant impacts with different impact geometries. In these simulations, the impactor is accreted by the target (Uranus/Neptune) and has a typically mass  between 1 and 3 M$_\oplus$.  Such giant impacts are not only required to explain the final planetary mass, but also for enriching the envelope metallicities. 
As a result, giant impacts on Uranus and Neptune seem to be a likely scenario that is consistent with our formation models. The fact that in many simulations the planets can accrete significant amounts of H-He at early stages also has implications for our understanding of  gaseous exoplanets with intermediate masses. Our models suggest that planets with smaller masses than those of Uranus and Neptune are likely to have higher H-He enrichment than those of Uranus and Neptune, as indicated by the large populations of mini-Neptunes \citep[e.g.,][]{otegi19}.

\section{Discussion}
While our study represents an advancement in our understanding of the formation of the outermost planets in the solar system, it is clear that further investigations are required. 
First, we do not find a given set of model parameters that can reproduce the formation of both Uranus and Neptune exactly. 
This could be resolved, for example, if the metallicities at 20 and 30 AU were different due to ice-lines between Uranus and Neptune  \citep{Powell17} that could altered the metallicity or the composition of the pebbles. Alternatively, it is possible that only one of the planets underwent a giant impact. It is clear that this topic should be investigated further in future studies. 
Second, our models assume isolated planets and do not include the interaction between Uranus and Neptune as they grow.
The pebble accretion rate is obtained under the assumption of an unperturbed disk where dust coagulates and drifts throughout the  protoplanetary disk \citep{Lambrechts14}.
However, the presence of another planet is expected to alter the disk structure and to opening gaps. This in turn can affect the pebble flux exerted at different orbital locations and therefore the planetary growth \citep{morbidelli20}. We therefore suggest that future studies should account for the simultaneous/consequent  formation of Uranus and Neptune. 
Third, here we used analytical prescriptions for the accretion rate of pebbles. This is of course an approximation and 2D/3D simulations of the interaction between the proto-planet and the disk are clearly required. 
Numerical studies suggest that pebble accretion may be halted prior to the disk's dissipation by the formation of a pressure bump \citep[e.g.,][]{Bitsch15,Bitsch18}, with important consequences for their further evolution \citep{Chen2020}.
Also, the heavy elements in this study are represented by pure-water. While pebbles are expected to be icy, they are likely to consist of various elements and future studies should implement EoS and opacity  calculations of various mixtures. This is important since different compositions lead to different formation paths (see Appendix~\ref{composition}). Accounting for various species for the heavy elements can also be used for providing  predictions on the planetary composition and its depth dependence. 

Finally, although our study focuses on the {\it in situ} formation of Uranus and Neptune it also has implications for the formation of intermediate-mass exoplanets.  
Since proto-planetary disks are expected to be diverse in metallicity, viscosity, and dissipation time, also  the resulting planets are expected to differ in final mass and composition. This explains the large diversity that we observe in the range of intermediate-mass planets. 
However, the disk's condition and the pebble accretion scenario may be  different for planets located at smaller orbital distances. 
Accretion-limiting processes such as entropy advection \citep[e.g.]{Ormel15,Moldenhauer21} might operate in the inner, but not in the outer disk where the local conditions are different. 

\section{Summary and Conclusions}

\label{conclusion}
Understanding the formation of Uranus and Neptune is crucial for understanding the origin of our solar system, as well as the formation of intermediate-mass exoplanets.  In this work we applied the giant planet formation model presented by VH20 to investigate the potential {\it in-situ} formation of Uranus and Neptune via pebble accretion.
We simulated the formation of Uranus and Neptune assuming different disk metallicities and viscosities.
\par

We find that Uranus and Neptune could have formed {\it in-situ} but that in many cases giant impacts are needed to explain the final mass and composition of the planets. The required mass of the impactor (or multiple impactors) depends on the model assumptions, such as the composition of the pebbles and the planet's opacity (see Appendix~\ref{composition} for discussion). The fact that a few Earth masses of heavy elements need to be accreted by Uranus and/or Neptune post-formation is consistent with the idea that Uranus and/or Neptune suffered giant impacts post-formation  \citep[e.g.,][]{Podolak2012, Reinhardt19}. 
We show that the primordial heavy-element distribution of Uranus and
Neptune is gradual implying that both planets have complex internal structure with composition gradients. 
Our main conclusions can be summarized as follows:
\begin{itemize}
    \item Pebble accretion, when accounting for the H-He accretion rate correctly, can lead to the formation of Uranus and Neptune at their current radial distances. The forming planets accrete H-He in proportions that are consistent with the estimated H-He mass from structure models, and the formation timescale is comprable to the expected lifetime of the solar nebula ($\sim$3 Myrs).
    \item In many cases the formed planets need to accrete between 1 to 3 M$_\oplus$ of heavy-element post-formation in order to reach their final masses and inferred compositions. This is consistent with the giant impact scenario which seems to be needed to explain several properties of Uranus and Neptune \citep[e.g.,][]{Reinhardt19}. 
    \item The primordial internal profiles can be roughly divided in three layers characterized by a different heavy-element mass fractions. 
    We show that the deep interiors of Uranus and Neptune are characterized by composition gradients. The planets have no distinct layers or well-defined core-envelope boundary. The primordial heavy-element profile reflects the ratio between the solid and gas accretion rates \citep{Helled17, Valletta20}.
\end{itemize}

In this paper we showed that Uranus and Neptune could have formed at their current  locations and that intermediate mass planets with H-He envelope is a   common output of planet formation via pebble accretion which is in agreement with exoplanetary data \citep[e.g.,][]{otegi19}. 
Whether Uranus and/or Neptune experienced giant impacts remains to be determined. 

It is clear that in order to better understand  the origin and evolution of Uranus and Neptune space missions dedicated to their exploration as well as progress in theory are required \citep[e.g.,][]{Helled19,Helled2020, 2021ExA...tmp...78F, Guillot2021}. Such efforts are now ongoing, and we hope that future measurements and theoretical studies will reveal the true nature of this unique planetary class. 

\section*{Acknowledgements}
We thank an anonymous referee for valuable comments. We also thank Peter Bodenheimer and Sho Shibata for valuable comments. We acknowledge support from the Swiss National Science Foundation (SNSF) under grant \texttt{\detokenize{200020_188460}}.

\bibliography{bibliography}

\begin{thebibliography}{}

\bibitem[\protect\astroncite{Bitsch et~al.}{2018}]{Bitsch18}
Bitsch, B., Morbidelli, A., Johansen, A., Lega, E., Lambrechts, M., and Crida,
  A. (2018).
\newblock The pebble isolation mass --- scaling law and implications for the
  formation of super-earths and gas giants.
\newblock {\em Astronomy \& Astrophysics}, 612.

\bibitem[\protect\astroncite{{Bitsch, Bertram} et~al.}{2015}]{Bitsch15}
{Bitsch, Bertram}, {Lambrechts, Michiel}, and {Johansen, Anders} (2015).
\newblock The growth of planets by pebble accretion in evolving protoplanetary
  discs.
\newblock {\em A\&A}, 582:A112.

\bibitem[\protect\astroncite{{Brouwers} et~al.}{2017}]{Brouwers2017}
{Brouwers}, {Vazan}, and {Ormel} (2017).

\bibitem[\protect\astroncite{{Brouwers} et~al.}{2021}]{Brouwers21}
{Brouwers}, M.~G., {Ormel}, C.~W., {Bonsor}, A., and {Vazan}, A. (2021).
\newblock {How planets grow by pebble accretion IV: Envelope opacity trends
  from sedimenting dust and pebbles}.
\newblock {\em arXiv e-prints}, page arXiv:2106.03848.

\bibitem[\protect\astroncite{Chachan et~al.}{2021}]{Chachan21}
Chachan, Y., Lee, E.~J., and Knutson, H.~A. (2021).
\newblock Radial gradients in dust-to-gas ratio lead to preferred region for
  giant planet formation.
\newblock {\em The Astrophysical Journal}, 919(1):63.

\bibitem[\protect\astroncite{{Chau} et~al.}{2020}]{chau20}
{Chau}, A., {Reinhardt}, C., {Izidoro}, A., {Stadel}, J., and {Helled}, R.
  (2020).
\newblock {Could Uranus and Neptune form by collisions of planetary embryos?}
\newblock In {\em AGU Fall Meeting Abstracts}, volume 2020, pages P066--0003.

\bibitem[\protect\astroncite{Chen et~al.}{2020}]{Chen2020}
Chen, Y.-X., Zhang, X., Li, Y.-P., Li, H., and Lin, D. N.~C. (2020).
\newblock Retention of long-period gas giant planets: Type {II} migration
  revisited.
\newblock 900(1):44.

\bibitem[\protect\astroncite{{Fletcher} et~al.}{2021}]{2021ExA...tmp...78F}
{Fletcher}, L.~N., {Helled}, R., {Roussos}, E., {Jones}, G., {Charnoz}, S.,
  {Andr{\'e}}, N., {Andrews}, D., {Bannister}, M., {Bunce}, E., {Cavali{\'e}},
  T., {Ferri}, F., {Fortney}, J., {Grassi}, D., {Griton}, L., {Hartogh}, P.,
  {Hueso}, R., {Kaspi}, Y., {Lamy}, L., {Masters}, A., {Melin}, H., {Moses},
  J., {Mousis}, O., {Nettleman}, N., {Plainaki}, C., {Schmidt}, J., {Simon},
  A., {Tobie}, G., {Tortora}, P., {Tosi}, F., and {Turrini}, D. (2021).
\newblock {Ice giant system exploration within ESA's Voyage 2050}.
\newblock {\em Experimental Astronomy}.

\bibitem[\protect\astroncite{Fortney and Nettelmann}{2010}]{Fortney2010}
Fortney, J.~J. and Nettelmann, N. (2010).
\newblock 152(1):423--447.

\bibitem[\protect\astroncite{Freedman et~al.}{2014}]{Freedman2014}
Freedman, R.~S., Lustig-Yaeger, J., Fortney, J.~J., Lupu, R.~E., Marley, M.~S.,
  and Lodders, K. (2014).
\newblock {\em The Astrophysical Journal Supplement Series}, 214(2):25.

\bibitem[\protect\astroncite{Goldreich et~al.}{2004}]{Goldreich04}
Goldreich, P., Lithwick, Y., and Sari, R. (2004).
\newblock Planet formation by coagulation: A focus on uranus and neptune.
\newblock {\em Annual Review of Astronomy and Astrophysics}, 42(1):549--601.

\bibitem[\protect\astroncite{{Guillot} et~al.}{2021}]{Guillot2021}
{Guillot}, T., {Fortney}, J., {Rauscher}, E., {Marley}, M.~S., {Parmentier},
  V., {Line}, M., {Wakeford}, H., {Kaspi}, Y., {Helled}, R., {Ikoma}, M.,
  {Knutson}, H., {Menou}, K., {Valencia}, D., {Durante}, D., {Ida}, S.,
  {Bolton}, S.~J., {Li}, C., {Stevenson}, K.~B., {Bean}, J., {Cowan}, N.~B.,
  {Hofstadter}, M.~D., {Hueso Alonso}, R., {Leconte}, J., {Li}, L.,
  {Mordasini}, C., {Mousis}, O., {Nettelmann}, N., {Soderlund}, K., and {Wong},
  M. (2021).
\newblock {Keys of a Mission to Uranus or Neptune, the Closest Ice Giants}.
\newblock In {\em Bulletin of the American Astronomical Society}, volume~53,
  page 244.

\bibitem[\protect\astroncite{Hayashi}{1981}]{Hayashi1981}
Hayashi, C. (1981).
\newblock {\em Progress of Theoretical Physics Supplement}, 70:35--53.

\bibitem[\protect\astroncite{Helled et~al.}{2010}]{Helled10}
Helled, R., Anderson, J.~D., Podolak, M., and Schubert, G. (2010).
\newblock {\em The Astrophysical Journal}, 726(1):15.

\bibitem[\protect\astroncite{{Helled} et~al.}{2011}]{Helled11}
{Helled}, R., {Anderson}, J.~D., {Podolak}, M., and {Schubert}, G. (2011).
\newblock {\em \apj}, 726:15.

\bibitem[\protect\astroncite{Helled and Bodenheimer}{2014}]{Helled14}
Helled, R. and Bodenheimer, P. (2014).
\newblock {\em The Astrophysical Journal}, 789(1):69.

\bibitem[\protect\astroncite{{Helled} and {Fortney}}{2020}]{Helled2020}
{Helled}, R. and {Fortney}, J.~J. (2020).
\newblock {The interiors of Uranus and Neptune: current understanding and open
  questions}.
\newblock {\em Philosophical Transactions of the Royal Society of London Series
  A}, 378(2187):20190474.

\bibitem[\protect\astroncite{Helled et~al.}{2020}]{Helled19}
Helled, R., Nettelmann, N., and Guillot, T. (2020).
\newblock Uranus and neptune: Origin, evolution and internal structure.
\newblock {\em Space Science Reviews}, 216(3):38.

\bibitem[\protect\astroncite{Helled and Stevenson}{2017}]{Helled17}
Helled, R. and Stevenson, D. (2017).
\newblock The fuzziness of giant planets' cores.
\newblock {\em The Astrophysical Journal}, 840(1):L4.

\bibitem[\protect\astroncite{Ida and Lin}{2004}]{ida04}
Ida, S. and Lin, D. N.~C. (2004).
\newblock Toward a deterministic model of planetary formation. i. a desert in
  the mass and semimajor axis distributions of extrasolar planets.
\newblock {\em The Astrophysical Journal}, 604(1):388--413.

\bibitem[\protect\astroncite{{Lambrechts} and {Johansen}}{2014}]{Lambrechts14}
{Lambrechts} and {Johansen} (2014).
\newblock {\em A\&A}, 572:A107.

\bibitem[\protect\astroncite{{Lambrechts} et~al.}{2014}]{lambrechts15}
{Lambrechts}, {Johansen, A.}, and {Morbidelli, A.} (2014).
\newblock Separating gas-giant and ice-giant planets by halting pebble
  accretion.
\newblock {\em A\&A}, 572:A35.

\bibitem[\protect\astroncite{Lissauer et~al.}{2009}]{lissauer2009}
Lissauer, J.~J., Hubickyj, O., D'Angelo, G., and Bodenheimer, P. (2009).
\newblock {\em Icarus}, 199(2):338 -- 350.

\bibitem[\protect\astroncite{Lozovsky et~al.}{2017}]{Lozovsky2017}
Lozovsky, M., Helled, R., Rosenberg, E.~D., and Bodenheimer, P. (2017).
\newblock {\em The Astrophysical Journal}, 836(2):227.

\bibitem[\protect\astroncite{{Moldenhauer} et~al.}{2021}]{Moldenhauer21}
{Moldenhauer}, {Kuiper, R.}, {Kley, W.}, and {Ormel, C. W.} (2021).
\newblock Steady state by recycling prevents premature collapse of
  protoplanetary atmospheres.
\newblock {\em A\&A}, 646:L11.

\bibitem[\protect\astroncite{Morbidelli et~al.}{2015}]{morbidelli15}
Morbidelli, A., Lambrechts, M., Jacobson, S., and Bitsch, B. (2015).
\newblock {\em Icarus}, 258:418 -- 429.

\bibitem[\protect\astroncite{{Morbidelli, A.}}{2020}]{morbidelli20}
{Morbidelli, A.} (2020).
\newblock Planet formation by pebble accretion in ringed disks.
\newblock {\em A\&A}, 638:A1.

\bibitem[\protect\astroncite{{Mordasini} et~al.}{2014}]{Mordasini14}
{Mordasini}, {Klahr, H.}, {Alibert, Y.}, {Miller, N.}, and {Henning, T.}
  (2014).
\newblock {\em A\&A}, 566:A141.

\bibitem[\protect\astroncite{Movshovitz and Podolak}{2008}]{Movshovitz08}
Movshovitz, N. and Podolak, M. (2008).
\newblock The opacity of grains in protoplanetary atmospheres.
\newblock {\em Icarus}, 194:368--378.

\bibitem[\protect\astroncite{{M\"uller} et~al.}{2020}]{Muller20}
{M\"uller}, {Helled, Ravit}, and {Cumming, Andrew} (2020).
\newblock The challenge of forming a fuzzy core in jupiter.
\newblock {\em A\&A}, 638:A121.

\bibitem[\protect\astroncite{Nettelmann et~al.}{2013}]{Nettelman13}
Nettelmann, N., Helled, R., Fortney, J., and Redmer, R. (2013).
\newblock New indication for a dichotomy in the interior structure of uranus
  and neptune from the application of modified shape and rotation data.
\newblock {\em Planetary and Space Science}, 77:143--151.
\newblock Surfaces, atmospheres and magnetospheres of the outer planets and
  their satellites and ring systems: Part VIII.

\bibitem[\protect\astroncite{{Nettelmann} et~al.}{2013}]{Nettelman2013}
{Nettelmann}, N., {Helled}, R., {Fortney}, J.~J., and {Redmer}, R. (2013).
\newblock {\em \planss}, 77:143--151.

\bibitem[\protect\astroncite{{Ormel} et~al.}{2021}]{Ormel21}
{Ormel}, {Vazan}, and {Brouwers} (2021).
\newblock How planets grow by pebble accretion - iii. emergence of an interior
  composition gradient.
\newblock {\em A\&A}, 647:A175.

\bibitem[\protect\astroncite{Ormel}{2014}]{Ormel14}
Ormel, C. (2014).
\newblock {AN} {ATMOSPHERIC} {STRUCTURE} {EQUATION} {FOR} {GRAIN} {GROWTH}.
\newblock {\em The Astrophysical Journal}, 789(1):L18.

\bibitem[\protect\astroncite{Ormel et~al.}{2015}]{Ormel15}
Ormel, C.~W., Shi, J.-M., and Kuiper, R. (2015).
\newblock {Hydrodynamics of embedded planets’ first atmospheres – II. A
  rapid recycling of atmospheric gas}.
\newblock {\em Monthly Notices of the Royal Astronomical Society},
  447(4):3512--3525.

\bibitem[\protect\astroncite{Otegi et~al.}{2019}]{otegi19}
Otegi, J., Bouchy, F., and Helled, R. (2019).

\bibitem[\protect\astroncite{Paxton et~al.}{2010}]{Paxton2010}
Paxton, B., Bildsten, L., Dotter, A., Herwig, F., Lesaffre, P., and Timmes, F.
  (2010).
\newblock {\em The Astrophysical Journal Supplement Series}, 192(1):3.

\bibitem[\protect\astroncite{Paxton et~al.}{2013}]{Paxton2013}
Paxton, B., Cantiello, M., Arras, P., Bildsten, L., Brown, E.~F., Dotter, A.,
  Mankovich, C., Montgomery, M.~H., Stello, D., Timmes, F.~X., and Townsend, R.
  (2013).
\newblock {\em The Astrophysical Journal Supplement Series}, 208(1):4.

\bibitem[\protect\astroncite{Paxton et~al.}{2015}]{Paxton2015}
Paxton, B., Marchant, P., Schwab, J., Bauer, E.~B., Bildsten, L., Cantiello,
  M., Dessart, L., Farmer, R., Hu, H., Langer, N., Townsend, R. H.~D.,
  Townsley, D.~M., and Timmes, F.~X. (2015).
\newblock {\em The Astrophysical Journal Supplement Series}, 220(1):15.

\bibitem[\protect\astroncite{Paxton et~al.}{2018}]{Paxton2018}
Paxton, B., Schwab, J., Bauer, E.~B., Bildsten, L., Blinnikov, S., Duffell, P.,
  Farmer, R., Goldberg, J.~A., Marchant, P., Sorokina, E., Thoul, A., Townsend,
  R. H.~D., and Timmes, F.~X. (2018).
\newblock {\em The Astrophysical Journal Supplement Series}, 234(2):34.

\bibitem[\protect\astroncite{{Paxton} et~al.}{2019}]{Paxton2019}
{Paxton}, B., {Smolec}, R., {Schwab}, J., {Gautschy}, A., {Bildsten}, L.,
  {Cantiello}, M., {Dotter}, A., {Farmer}, R., {Goldberg}, J.~A., and {Jermyn},
  A.~S. (2019).
\newblock {\em arXiv e-prints}, page arXiv:1903.01426.

\bibitem[\protect\astroncite{Pfalzner et~al.}{2014}]{Pfalzner14}
Pfalzner, S., Steinhausen, M., and Menten, K. (2014).
\newblock {\em The Astrophysical Journal}, 793(2):L34.

\bibitem[\protect\astroncite{Piso and Youdin}{2014}]{Piso14}
Piso, A.-M.~A. and Youdin, A.~N. (2014).
\newblock {\em The Astrophysical Journal}, 786(1):21.

\bibitem[\protect\astroncite{{Podolak} and {Helled}}{2012}]{Podolak2012}
{Podolak}, M. and {Helled}, R. (2012).
\newblock {What Do We Really Know about Uranus and Neptune?}
\newblock {\em \apjl}, 759(2):L32.

\bibitem[\protect\astroncite{Pollack et~al.}{1996}]{POLLACK1996}
Pollack, J.~B., Hubickyj, O., Bodenheimer, P., Lissauer, J.~J., Podolak, M.,
  and Greenzweig, Y. (1996).
\newblock {\em Icarus}, 124(1):62 -- 85.

\bibitem[\protect\astroncite{Powell et~al.}{2017}]{Powell17}
Powell, D., Murray-Clay, R., and Schlichting, H.~E. (2017).
\newblock Using ice and dust lines to constrain the surface densities of
  protoplanetary disks.
\newblock {\em The Astrophysical Journal}, 840(2):93.

\bibitem[\protect\astroncite{{Rafikov}}{2011}]{Rafikov2011}
{Rafikov}, R.~R. (2011).
\newblock {\em \apj}, 727:86.

\bibitem[\protect\astroncite{Reinhardt et~al.}{2019}]{Reinhardt19}
Reinhardt, C., Chau, A., Stadel, J., and Helled, R. (2019).
\newblock {Bifurcation in the history of Uranus and Neptune: the role of giant
  impacts}.
\newblock {\em Monthly Notices of the Royal Astronomical Society},
  492(4):5336--5353.

\bibitem[\protect\astroncite{Reinhardt et~al.}{2020}]{Reinhardt20}
Reinhardt, C., Chau, A., Stadel, J., and Helled, R. (2020).
\newblock Bifurcation in the history of uranus and neptune: the role of giant
  impacts.
\newblock {\em Monthly Notices of the Royal Astronomical Society},
  492:5336--5353.

\bibitem[\protect\astroncite{{Ribas} et~al.}{2015}]{Ribas15}
{Ribas}, {Bouy, Herv\'e}, and {Mer\'{\i}n, Bruno} (2015).
\newblock {\em A\&A}, 576:A52.

\bibitem[\protect\astroncite{{Safronov}}{1972}]{Safronov1969}
{Safronov}, V.~S. (1972).

\bibitem[\protect\astroncite{{Shibata} et~al.}{2020}]{Shibata20}
{Shibata}, {Helled}, and {Ikoma} (2020).
\newblock The origin of the high metallicity of close-in giant exoplanets -
  combined effects of resonant and aerodynamic shepherding.
\newblock {\em A\&A}, 633:A33.

\bibitem[\protect\astroncite{Shibata and Ikoma}{2019}]{Shibata19}
Shibata, S. and Ikoma, M. (2019).
\newblock {\em Monthly Notices of the Royal Astronomical Society},
  487(4):4510--4524.

\bibitem[\protect\astroncite{Stevenson}{1982}]{stevenson82}
Stevenson, D. (1982).
\newblock Formation of the giant planets.
\newblock {\em Planetary and Space Science}, 30(8):755--764.

\bibitem[\protect\astroncite{Thommes et~al.}{1999}]{thommes99}
Thommes, E.~W., Duncan, M.~J., and Levison, H.~F. (1999).
\newblock {\em Nature}, 402(6762):635--638.

\bibitem[\protect\astroncite{Tsiganis et~al.}{2005}]{tsiganis05}
Tsiganis, K., Gomes, R., Morbidelli, A., and Levison, H.~F. (2005).
\newblock {\em Nature}, 435(7041):459--461.

\bibitem[\protect\astroncite{Valencia et~al.}{2013}]{Valencia2013}
Valencia, D., Guillot, T., Parmentier, V., and Freedman, R.~S. (2013).
\newblock {\em The Astrophysical Journal}, 775(1):10.

\bibitem[\protect\astroncite{{Valletta} and {Helled}}{2020}]{Valletta20}
{Valletta} and {Helled} (2020).
\newblock {\em The Astrophysical Journal}, 900(2):133.

\bibitem[\protect\astroncite{Valletta and Helled}{2019}]{Valletta19}
Valletta, C. and Helled, R. (2019).
\newblock {\em The Astrophysical Journal}, 871(1):127.

\bibitem[\protect\astroncite{{Venturini} et~al.}{2016}]{Venturini16}
{Venturini}, {Alibert, Yann}, and {Benz, Willy} (2016).
\newblock {\em A\& A}, 596:A90.

\bibitem[\protect\astroncite{Venturini and Helled}{2017}]{Venturini17}
Venturini, J. and Helled, R. (2017).
\newblock {\em The Astrophysical Journal}, 848.

\end{thebibliography}

\appendix

\section{The importance of opacity and composition of the pebbles}
\label{composition}
As shown in previous studies,  the envelope's opacity significantly affects the H-He accretion rate \citep[e.g.,][]{Brouwers21,Brouwers2017,Mordasini14,Movshovitz08}.
The opacity that we used in this study corresponds to the sum of the gas and dust opacity.
Determining the grain opacity, however, is challenging since it depends on many parameters, such as the number of grains present in a layer, their composition, shape, and size distribution. For simplicity, the dust grain opacity is simply calculated using the prescription of \cite{Valencia2013}. 

The dust opacity derived by \cite{Valencia2013} is significantly higher than suggested by other studies \citep{Mordasini14,Ormel14,Brouwers21}. 
However, the grain opacity is expected to be high at early phases of the planetary growth  when the protoplanet accretes pebbles due to efficient pebble fragmentation \citep{Brouwers21,Chachan21}, pebble ablation \citep[e.g.,][]{Valletta20} and bouncing collisions between accreted pebbles \citep{Ormel21}. 

\begin{figure}[h]
	\centering
	\includegraphics[width=0.7\linewidth]{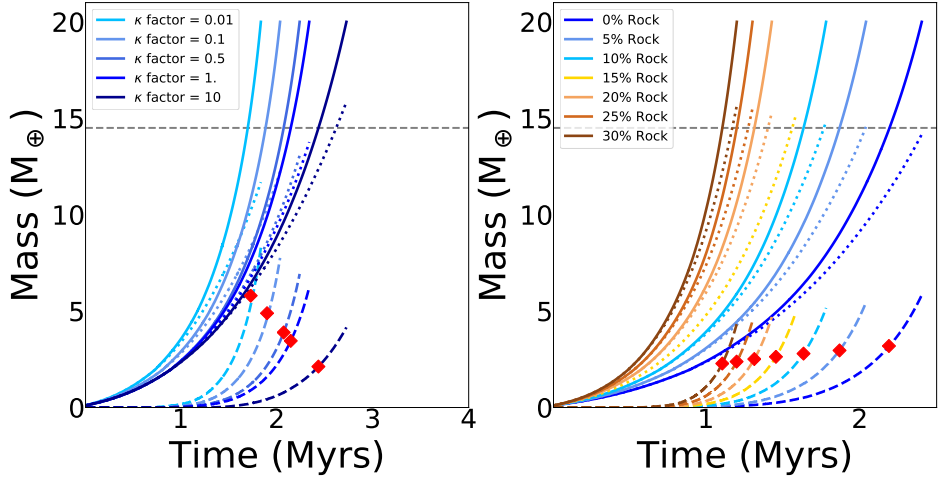}
	\caption{\textbf{Left:} Planetary mass vs.~time for various assumed opacities. The solid, dashed, and dotted lines indicate the total planetary mass, the H-He mass and the mass of heavy elements, respectively. Different colours represent different opacity factors. 
	\textbf{Right:} Planetary mass as a function of time with the different colours indicating different assumed pebble densities (i.e., composition). In both panels the red dots indicate the time at which the planet's mass is 14.5 M$_\oplus$.
	}
	\label{Uranus-Neptune3}
\end{figure}

Here in order to demonstrate the importance of the dust opacity on the planetary growth we multiply the dust opacity by a fixed factor. The results are shown in Figure \ref{Uranus-Neptune3} where different line colours represent cases with the opacity is multiplied by different factors. 
It is clear from the figure that the planetary growth strongly depends on the dust opacity: higher dust opacity result in less efficient accretion of gas. 

The composition of the accreted pebbles is also important since it can affect the solid accretion rate, due to the modification of the Stokes number. 
Pebbles with higher densities lead to higher solid accretion rates and therefore shorter formation timescales. 
The stokes number depends on the pebble density according to Eq.~6 of \cite{Lambrechts14}. In turn, the solid accretion rate is affected by the Stokes number as can be seen in Eq.~\ref{3dsacc}.

In the right panel plot of Figure \ref{Uranus-Neptune3} we show the planetary growth for different assumed pebble  composition.
Different colours represent different assumed pebble's density. Rock is represented by SiO$_2$, with a density of 2.7 g cm$^{-3}$.
As expected, higher rock fractions lead to higher solid accretion rates. 
It should be noted that for simplicity in these calculations the pebble composition is not considered in the equation of state and opacity calculations.

\section{Dependence on the metallicity and viscosity of the disk}
\label{parameters}
In this section, we investigate a larger parameter space of disk's metallicity and viscosity. 

The range of parameters is summarized in Table~\ref{table1}, where the names of the runs names reflect the assumed parameters: the first letter of the run's name of U or N corresponds to the formation of Uranus or Neptune, respectively. 
The time at which the simulation is terminated essentially corresponds to the required disk's dissipation time, which is indicated in the fourth column of the table. 
The second and third columns in Table~\ref{table1} represent the disk's metallicity in solar units (we use Z$_\odot = 0.012$) and the disk's viscosity, respectively. 
The metallicity and viscosity of the disk are also indicated by the run name. For example, U130 indicates a metallicity equals to 1.30 Z$_\odot$, while "1-3" represents a disk's viscosity of $1 \times 10^{-3}$. 
The fourth column indicates the missing heavy-element mass that needs to be accreted after the dissipation of the solar nebula. If $\Delta$ M = 0 the planet have reached the mass of Uranus/Neptune when the disk dissipates. 
In this case if a giant impact occurs after the disk's dissipation the mass of heavy elements in the planet can exceeds the amount inferred by observations.
The fifth column indicates the formation probability $f_{U,N}$ (Eq.~\ref{probability}) where the disk's dissipation time is assumed to obey a gaussian probability function, while the sixth ($f_{U,N}-V$) column shows the formation probability as computed via equation 2 of \cite{Venturini17}.

\begin{figure}[h]
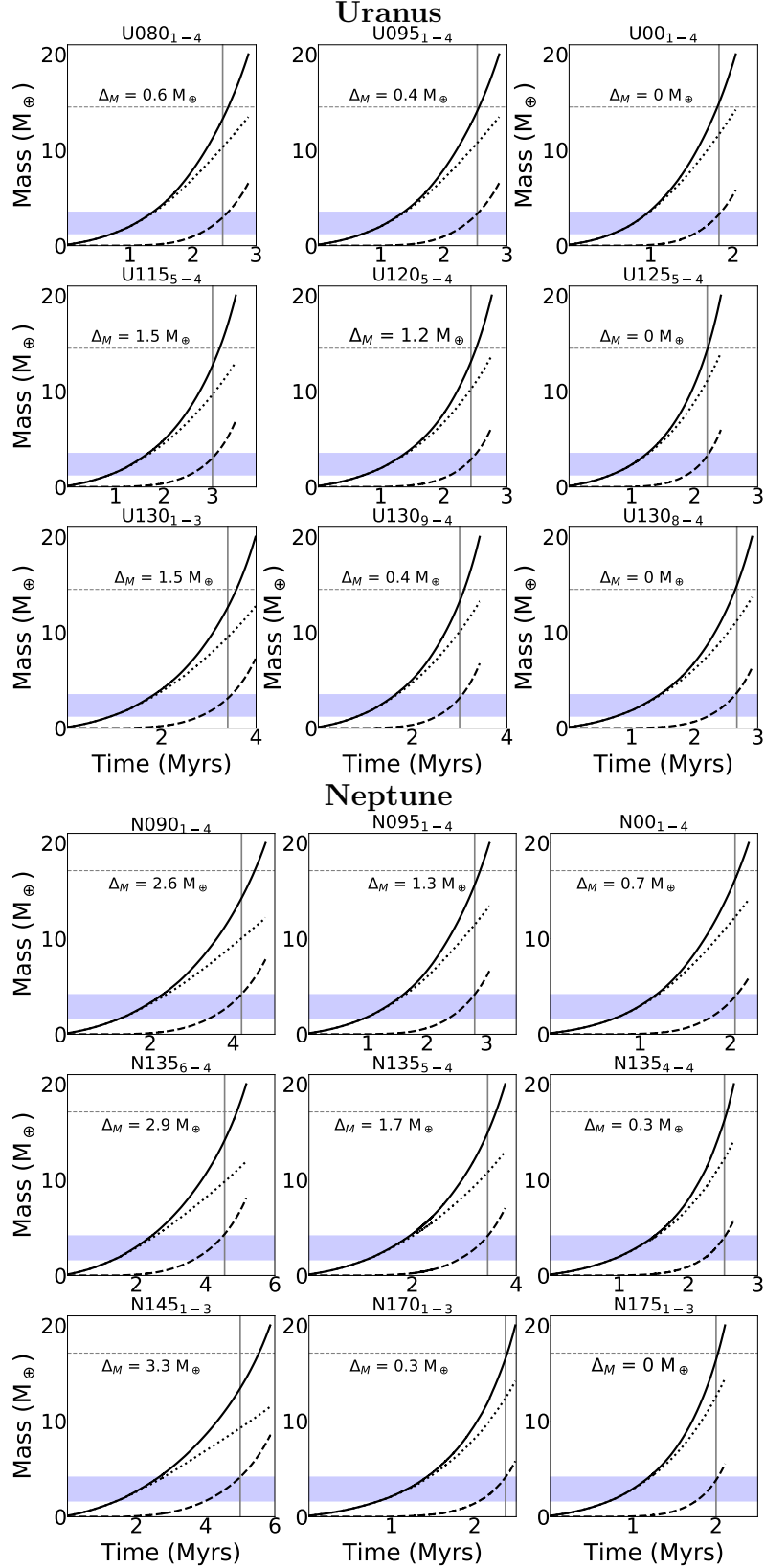

	\centering
	{\large \bf Uranus}\\
	\includegraphics[width=0.58\linewidth]{ParameterSpace.pdf}\\
	{\large \bf Neptune}\\
	\includegraphics[width=0.58\linewidth]{ParameterSpace2.pdf}
	\caption{Planetary mass vs.~time from various formation simulations of Uranus (top 9 panels) and Neptune (bottom 9 panels). The solid, dashed, and dotted lines represent the total planetary mass, the H-He mass, and the mass of heavy elements, respectively. 
	The grey horizontal lines indicate the masses of Uranus and Neptune, while the vertical grey line shows the requested disk's dissipation time.
	The blue region indicates the H-He in Uranus and Neptune as inferred from interior models. 
	The parameters of the different runs are listed in Table~\ref{table1}.
    The first two lines of both figures show the effect of changing the disk's metallicity keeping the other parameters fixed, while the third line of the two figures shows the effect of changing the disk's opacity, keeping the other parameters fixed.}
	
	\label{ParameterSpace}
\end{figure}

The first two runs indicated in Table~\ref{table1} (U145$_{1-3}$ and N145$_{5-4}$) are the ones presented in Figure~\ref{Uranus-Neptune}.
The top panels of Figure~\ref{Uranus-Neptune2} show runs U140$_{1-3}$ and N140$_{5-4}$ while in the bottom panels are shown runs U145$_{1.5}$ and N145$_{7.5-4}$. The other runs listed in Table~\ref{table1} are presented in Figure~\ref{ParameterSpace}.

The runs listed in Table~\ref{table1} include disk metallicities between 0.8 and 1.75 Z$_\odot$ and a disk viscosities between 10$^{-3}$ and 10$^{-4}$.
We find that for these assumed ranges, the inferred required dissipation time is in agreement with the expected lifetime of the solar nebula,  1.8 -- 4 Myrs. 
In many cases a couple M$_{\oplus}$ of heavy elements are missing when the disk dissipates, suggesting that Uranus and/or Neptune must accrete $\sim$1 -- 3 M$_\oplus$ of heavy elements post formation. This could be in the form of giant impacts, which is consistent with the idea that both Uranus and Neptune suffered giant impacts post-formation, that can explain their observed physical properties \citep[e.g.,][]{Podolak2012, Reinhardt19}.  
However, there is a sufficient number of cases in which the planets reach the measured mass of Uranus/Neptune by the end of the simulation and no further accretion is required.
Finally, Table~\ref{table1} also lists the formation probability for the different scenarios. We find that in general the formation probability as defined in \cite{Venturini17} is lower than our definition given in Eq.~\ref{probability}. However, for both definitions, the presented scenarios are plausible. The formation probability goes up to 35\% according to our definition and to 18\% according to the definition provided in \cite{Venturini17}.
This suggests that Uranus- and Neptune- like planet are a rather  common outcome of pebble accretion at large orbital distances.
Generally speaking, scenarios requesting giant impacts are found to be more probable. 
In the cases where extra mass of heavy elements are required to be accreted post-formation, the planetary growth is slower and the protoplanet spends more time in the blue region of the plots (the region with the correct mass of H-He). 

Figure~\ref{ParameterSpace} demonstrates that for given assumed parameters for the solar nebula we predict different formation paths for Uranus and Neptune as we don't find a setup which reproduced the observed properties of both planets.  
For example, in U00$_{1-4}$ and N00$_{1-4}$ the same set of parameters are assumed at 20 and 30 AU, however in this case we find that Uranus reaches its final mass and expected composition, while Neptune needs to accrete an extra 0.9 M$_\oplus$ of heavy elements.  
In the case of U095$_{1-4}$ and N095$_{1-4}$ both planets don't reach their final masses by the end of the simulations, and both Uranus and Neptune need to accrete $\sim$ 1 and 1.5 M$_\oplus$ of heavy elements post-formation. Interestingly, we find that in most of these cases we have intermediate mass planets with non-negligible H-He envelope, consistent with the observed exoplanet populations.  

\begin{table}[h!]
\begin{center}
\begin{tabular}{ccccccccc}\hline

Run name &Disk's metallicity (Z$_\odot$) &Disk's viscosity &Dissipation time (Myr) &$\Delta$ M (M$_\oplus$) &$f_{U,N}$&$f_{U,N}$-V\\ \hline

U145$_{1-3}$ & 1.45 Z$_\odot$ &   10$^{-3}$ & 2.1 & 0 M$_\oplus$  &9.5 \% &9.4 \% \\
N145$_{5-4}$ & 1.45 Z$_\odot$ &  5 $\times$ 10$^{-4}$ & 2.4 & 0 M$_\oplus$ &10.5 \% &8.7 \%\\
U140$_{1-3}$ & 1.40 Z$_\odot$ &   10$^{-3}$ & 2.5 & 0.8 M$_\oplus$  &14.5\% &10.4\%\\
N140$_{5-4}$ & 1.40 Z$_\odot$ &  5 $\times$ 10$^{-4}$ & 2.83 &1.3 M$_\oplus$ &18.3\% &10.3\%\\
U145$_{1.5}$ & 1.45 Z$_\odot$ &  1.5 $\times 10^{-3}$ & 3.5 & 2.4 M$_\oplus$  &35 \% &18.3 \%\\
N145$_{7.5-4}$ & 1.45 Z$_\odot$ &  7.5 $\times 10^{-4}$ & 4 & 2.6 M$_\oplus$  &25\% &11.3\% \\

U080$_{1-4}$ & 0.80 Z$_\odot$ &  $10^{-4}$ & 2.48 & 0.6 M$_\oplus$  &15.7\% & 11.3\%  \\
U095$_{1-4}$ & 0.90 Z$_\odot$ &  $10^{-4}$ & 2.5 & 0.4 M$_\oplus$  &16\% &11.2\%\\
U00$_{1-4}$ & 1.00 Z$_\odot$  &  $10^{-4}$ & 1.82 & 0 M$_\oplus$  &6.3\% & 10.6\%\\
U115$_{5-4}$ & 1.15 Z$_\odot$ &  5 $\times 10^{-4}$ & 3.1 & 1.5 M$_\oplus$  &24.5\% & 12.4\% \\
U120$_{5-4}$ & 1.20 Z$_\odot$ &  5 $\times 10^{-4}$ & 2.4 & 1.2 M$_\oplus$  &14.2\% & 10.5\% \\
U125$_{5-4}$ & 1.25 Z$_\odot$ &  5 $\times 10^{-4}$ & 2.2 & 0 M$_\oplus$  &10.0\% & 9.9\% \\
U130$_{1-3}$ & 1.30 Z$_\odot$ &  $10^{-3}$ & 3.4 & 1.5 M$_\oplus$  &27.5\% & 12.3\%\\
U130$_{9-4}$ & 1.30 Z$_\odot$ &  9 $\times 10^{-4}$ & 3 & 0.4 M$_\oplus$  &22.7\% & 11.4\%\\
U130$_{8-4}$ & 1.30 Z$_\odot$ &  8 $\times 10^{-4}$ & 2.7 & 0 M$_\oplus$  &16.8\% & 10.9\%\\

N090$_{1-4}$ & 0.90 Z$_\odot$ & $10^{-4}$ & 4.20 & 2.6$_\oplus$  &23\% &12.2\% \\
N095$_{1-4}$ & 0.95 Z$_\odot$ & $10^{-4}$ & 2.80 & 1.3 M$_\oplus$  &18.7\% &10.8\% \\
N00$_{1-4}$ & 1.00 Z$_\odot$  & $10^{-4}$ & 2.05 & 0.7 M$_\oplus$  &7.7\% &9.4\% \\
N135$_{6-4}$ & 1.35 Z$_\odot$ & 6 $\times 10^{-4}$ & 4.55 & 2.9 M$_\oplus$  &23\% &12.8\% \\
N135$_{5-4}$ & 1.35 Z$_\odot$ & 5 $\times 10^{-4}$ & 3.45 & 1.7 M$_\oplus$  & 24.5\% & 11.6\% \\
N135$_{4-4}$ & 1.35 Z$_\odot$ & 4 $\times 10^{-4}$ & 2.52 & 0.3 M$_\oplus$  &13.1\% &9.1\%\\
N145$_{1-3}$ & 1.45 Z$_\odot$ & $10^{-4}$ & 5 & 3.3 M$_\oplus$  &17.2\% &12.9\% \\
N170$_{1-3}$ & 1.70 Z$_\odot$ & $10^{-3}$ & 2.36 & 0.3 M$_\oplus$  &10.9\% &8.9\% \\
N175$_{1-3}$ & 1.75 Z$_\odot$ & $10^{-3}$ & 2 & 0 M$_\oplus$  &6.4\% &7.8\% \\

\hline

\end{tabular}
\end{center}

\caption{The various cases considered. In the run names the 'U' and the 'N' represent the formation of Uranus and Neptune, respectively. 
In the first case the orbital's location is 20 AU and we stop the simulation when the proto-planet has accreted 3 M$_\oplus$ of H-He, while in the second case the orbital's location is 30 AU and the simulations are stopped when the planet accretes 4 M$_\oplus$ of H-He. 
The column $\Delta$ M indicates the missing heavy-element mass when the disks assumed to dissipate. As we discuss above, this missing mass could be added via giant impacts or late planetesimal accretion. If $\Delta$ M = 0 it means that the planet have reached Uranus/Neptune mass by the end of the simulation and no further accretion is required. 
Finally, $f_{U,N}$ represent the probability of a given formation scenario as defined in Eq. 2.  }

\label{table1}
\end{table}

\begin{figure}[h]
	\centering
	\includegraphics[width=0.8\linewidth]{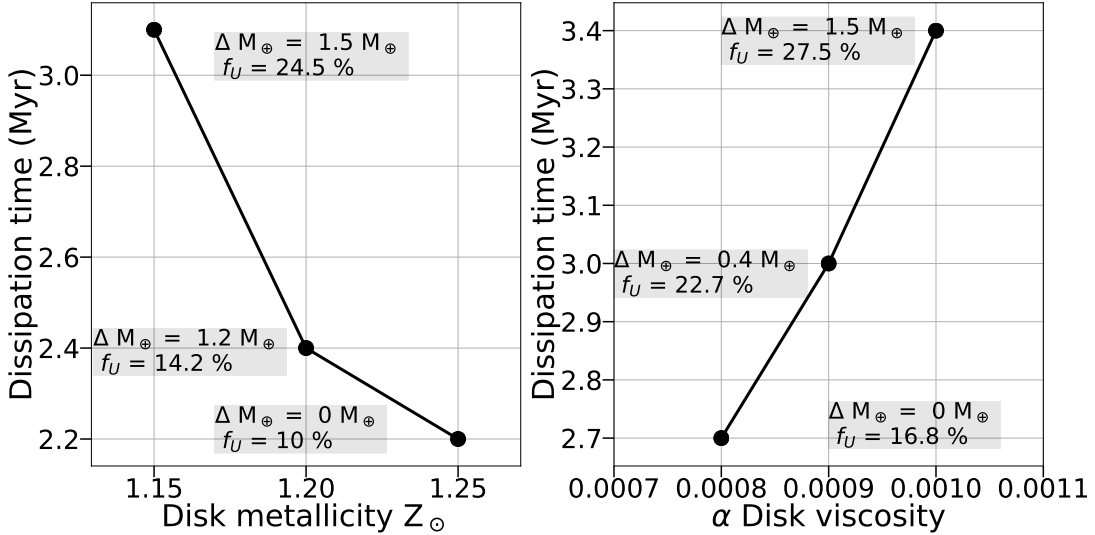}
	\caption{\textbf{Left:} Disk dissipation time vs>~disk's metallicity (in solar units). \textbf{Right:} Disk dissipation time vs.~disk's viscosity. The text boxes indicate $\Delta$ M and the formation probability.
	}
	\label{Metallicityviscosity}
\end{figure}

In Figure~\ref{Metallicityviscosity} we show in detail how the disk's dissipation time, the $\Delta$ M and the formation probability depend on the disk's metallicity and disk's viscosity. 
The left panel shows runs U115$_{5-4}$, U120$_{5-4}$ and U125$_{5-4}$. 
The various model parameters are kept fixed and we investigate the sensitivity of the results to the assumed disk's metallicity.
We find that the dissipation time, $\Delta$ M and the formation probability decrease with  increasing disk metallicity. 
The right panel plot shows runs U130$_{9-4}$, U130$_{8-4}$ and U130$_{1-3}$. Here, the  metallicity is fixed and we investigate how our results depend on the disk's viscosity. 
We find that the dissipation time $\Delta$ M and the formation probability increase with increasing viscosity.

\section{Internal composition of Uranus and Neptune}
\label{internalcomposition}

\begin{figure}
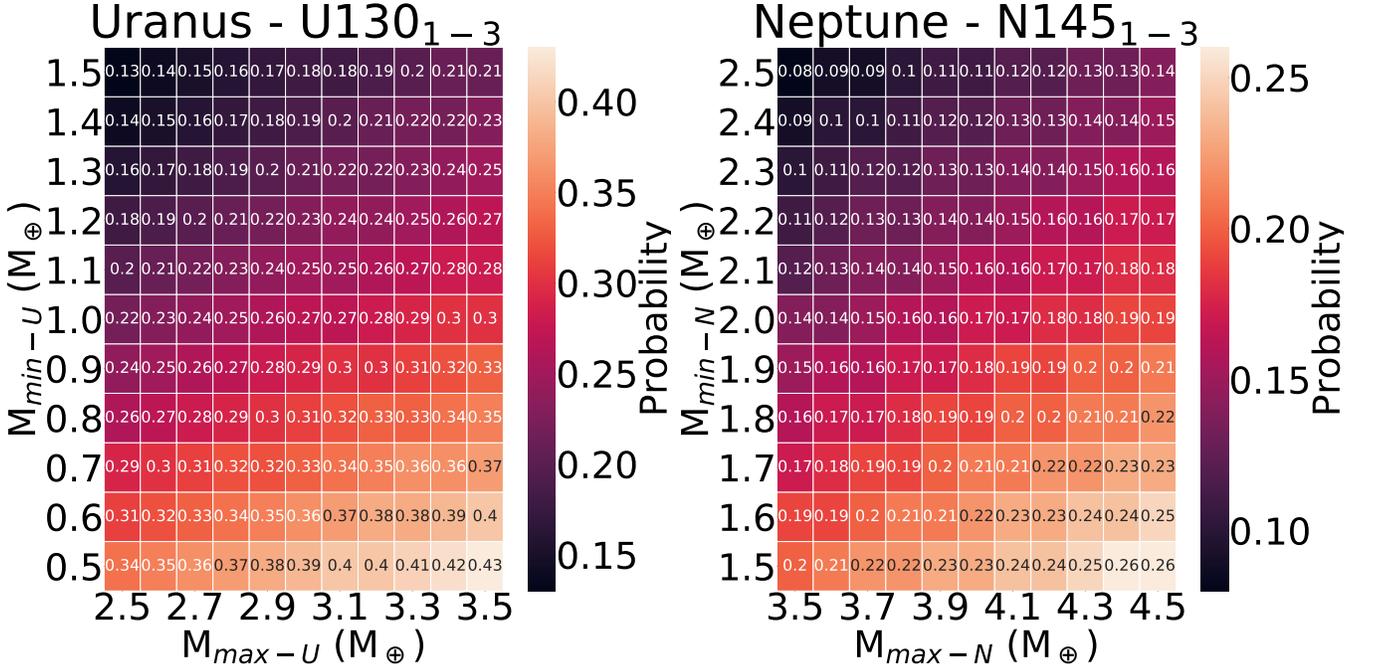

\label{minmaxlikelihood}
	   \centering
	    \includegraphics[width=0.49\linewidth]{Uranus.pdf}
		\includegraphics[width=0.49\linewidth]{Neptune.pdf}
	    \caption{Formation probability as defined via Eq.~\ref{probability} vs.~M$_{max-U(N)}$ and M$_{min-U(N)}$, which are represented by the x and y-axis, respectively. The left and right panels correspond to runs  U130$_{1-3}$ and N145$_{1-3}$, respectively.}
		\label{Fig2}
\end{figure}

The computed formation probability shown in the last two columns of Table~\ref{table1} depends on the assumed minimum and maximum mass of H-He  in Uranus and Neptune.
We assumed that M$_{min-U(N)}$ and M$_{max-U(N)}$ are 1.25 (1.64) M$_\oplus$ and M$_{max-U(N)}$ is 3.53 (4.15) M$_\oplus$.
In figure~\ref{minmaxlikelihood} we explore the sensitivity of the formation likelihood on M$_{min-U(N)}$ and M$_{max-U(N)}$.
In the left and right panel plot we show the formation probability of forming Uranus (represented by run U130$_{1-3}$) and Neptune (represented by run N145$_{1-3}$), respectively.
The x and y-axis indicate M$_{max-U(N)}$ and M$_{min-U(N)}$, respectively. The formation probability is indicated by a color scheme. 
We find that when assuming a low M$_{min-U(N)}$ and a large M$_{max-U(N)}$ the formation probability of Uranus and Neptune is relatively high.  
In particular, the left panel plot suggests  that assuming a maximum and a minimum H-He Uranus mass of 0.5 and 3.5 M$_\oplus$ results in a formation probability $\sim$ 40\%.
On the other hand, the formation probability becomes significantly smaller when assuming  low M$_{max-U(N)}$ and high M$_{min-U(N)}$.

\newpage
\section{Pebble accretion rate}
\label{saccrateapp}
The pebble (i.e., heavy-element) accretion rate implemented in this work is given by \citep{Lambrechts14}:

\begin{equation}
\label{saccrate}
\dot{M}_Z = 4.8 \times 10^{-6} \bigg(\frac{M_p}{M_\oplus}\bigg)^{2/3} \bigg(\frac{Z_0}{0.01}\bigg)^{25/18}  \bigg(\frac{M_\star}{M_\odot}\bigg)^{-11/36} \bigg(\frac{\beta}{500 g cm^{-2}}\bigg)  \bigg(\frac{r}{10 AU}\bigg)^{-5/12}  \bigg(\frac{t}{10^6 yr}\bigg)^{-5/18} M_\oplus yr^{-1},
\end{equation}
where $M_p$ and $M_\star$ represent the mass of the planet and the mass of the central star, respectively. $Z_0$ and $\beta$ indicate the disk metallicity and the gas surface density, respectively. Finally, $t$ and $r$ represent the protoplanet's age and the orbital location.
The pebble accretion rate transitions to the lower 3D regime when:
\begin{equation}
    \frac{\pi (\tau_f/0.1)^{1/3} R_H}{2 \sqrt{2\pi}}<H_{peb},
\end{equation}
where $H_{peb}$, $\tau_f$ and $R_{H}$ represent the pebble scale height, the Stokes number, and the Hill's radius, respectively. 
In this case, the pebble accretion rate is given by: \citep{morbidelli15}:
\begin{equation}
\label{3dsacc}
    \dot{M}_{Z,3d} = \dot{M}_Z \times \frac{\pi (\tau_f/0.1)^{1/3} R_H}{2 \sqrt{2\pi}H_{peb}},
\end{equation}
where $\dot{M}_Z$ represents the 2D solid accretion rate given in Eq.~\ref{saccrate}.

\section{Disk model}
\label{diskmodelapp}
Our baseline model is similar to the one presented in \cite{Lambrechts14} where it is assumed that the gas surface density of the disk obeys a simple power law:
\begin{equation}
    \Sigma_g = \Sigma_0 \bigg( \frac{r}{AU} \bigg)^{-1},
\end{equation}
where $\Sigma_0 = 500$ g cm$^{-2}$. 
For the temperature and pressure profiles we adopt a passively irradiated disk model \citep{Piso14} where:
\begin{equation}
    T(r) = 45 \bigg(\frac{a}{10 AU}\bigg)^{-3/7} K
\end{equation}
\begin{equation}
    P(r) = 6.9 \times 10^{-3} \bigg(\frac{a}{10 AU}\bigg)^{-45/14} \mathrm{dyn \; cm^{-2}}
\end{equation}
We use an analytical approximation to model the pebble accretion rate (Eq.~\ref{saccrate}). 
The mass of accreted pebbles depends on the local solid surface density at the planet's formation. As a result, the results of our simulations do not depend on the implemented disk model as long as different disk models predict a similar solid surface density at the planet's formation location. 
For example, in \cite{Venturini17} the authors followed the minimum mass solar nebula model \citep{Hayashi1981}; the gas surface density is given by:
\begin{equation}
\label{diskmodel}
    \Sigma_g = \Sigma_0 (r/AU)^{-3/2},
\end{equation}
where $\Sigma_0 = 1700$ g cm$^{-2}$. 
Assuming a metallicity of 0.018 (which is the baseline model in \cite{Venturini17}) the solid surface density predicted by Eq.~\ref{diskmodel} at 20 AU is similar to our disk model assuming a metallicity of 1 Z$_\odot$ which is the case shown in the top right plot in Figure~\ref{ParameterSpace}.

\end{document}